\documentclass[12pt]{article}
\usepackage{geometry}             
\usepackage{graphicx}
\usepackage{amssymb}
\usepackage{amsmath}
\usepackage{epstopdf}
\usepackage{comment}
\usepackage{cite}
\usepackage{abstract}

\usepackage[hyperindex=true,
          pdfstartview=FitH,
          bookmarksnumbered=true,
          bookmarksopen=true,
          citecolor=blue,
          linkcolor=blue,
          colorlinks=true,
          unicode]{hyperref}

\usepackage{makecell}

\begin{document}
\title{Black hole thermodynamics and heat engines in conformal gravity}
\author{Hao Xu, Yuan Sun and Liu Zhao\\
School of Physics, Nankai University,
Tianjin 300071, China\\
\em email:
\href{mailto: haoxu@mail.nankai.edu.cn}{haoxu@mail.nankai.edu.cn,}
\href{mailto: sunyuan14@mail.nankai.edu.cn}{sunyuan14@mail.nankai.edu.cn}\\
and \href{mailto: lzhao@nankai.edu.cn }{lzhao@nankai.edu.cn}
}

\date{}                             
\maketitle

\begin{abstract}
The extended phase space thermodynamics and heat engines for static spherically symmetric
black hole solutions of four dimensional conformal gravity are studied in detail.
It is argued that the equation of states (EOS) for such black holes is always branched, any
continuous thermodynamical process cannot drive the system from one branch of the EOS
into another branch. Meanwhile, the thermodynamical volume is bounded from above,
making the black holes always super-entropic in one branch and may also be super-entropic in
another branch in certain range of the temperature. The Carnot and Stirling heat engines
associated to such black holes are shown to be distinct from each other. For rectangular
heat engines, the efficiency always approaches zero when the rectangle becomes extremely
narrow, and given the highest and lowest working temperatures fixed, there is always a maximum
for the efficiency of such engines.
\end{abstract}

\section{Introduction}

Among various higher curvature gravity models in four dimensions, conformal gravity
defined by the action
\begin{align}
S=\alpha\int \mathrm{d}^4x \sqrt{-g}C^{\mu\nu\rho\sigma}
C_{\mu\nu\rho\sigma},
\label{action}
\end{align}
where $\alpha$ is a parameter with dimension [length]$^2$ (and will be set equal to 1
which effectively fixes the unit system to be used in this paper), is the single
distinguished model which exhibits conformal invariance and various other
exotic features. The model is often described as one that is only sensitive to angles but
not to distances due to the conformal invariance. However, since the conformal symmetry
is only the symmetry of the action but not of the metric tensor, and in particular,
conformal transformations may change the conformal infinities and thus the global
structure of the spacetime, the ``insensitive to distance'' claim is actually not sounded.
Moreover, the model contains ghost degrees of freedom in its perturbative mass spectrum,
making it a little difficult to be accepted as a realistic theory of gravitation.

Things get changed after Maldacena discovered \cite{Maldacena:2011mk}
that the on-shell action of conformal gravity
in AdS background can be made equal to that of Einstein gravity, provided one takes the
Neumann boundary condition. This fact, together with the features that conformal gravity
is power counting renormalizable and that it can fit the galaxy rotation curve well without
the introduction of dark matter \cite{Mannheim:1988dj,Mannheim:2005bfa,Mannheim:2010ti,Mannheim:2011ds}, make conformal gravity a very attractive alternative model
of gravity in four dimensions.

Because of the conformal symmetry of its action, conformal gravity allows much more solutions
than Einstein gravity. For instance, the Weyl rescaling of a static spherically symmetric
solution with the Weyl factor depending only on the radial coordinate remains a static
spherically symmetric solution, thus the Birkhoff theorem can only restrict the static
spherically symmetric solutions to a conformal class. It was proven in \cite{Riegert:1984zz}
that any static spherically symmetric solution of conformal gravity is a member in the
conformal class of the following metric,
\begin{align}
  &\mathrm{d} s^2=-f(r)\mathrm{d}t^2+\frac{\mathrm{d}r^2}{f(r)}+r^2\mathrm{d}
  \Omega^2, \label{metric}
\end{align}
where $\mathrm{d}\Omega^2$ is the line element of a 2-dimensional sphere and
\begin{align}
  f(r)=c_0+\frac{d}{r}+c_1 r-\frac{1}{3}\Lambda r^2.
\end{align}
There are four different integration constant $ c_0, c_1, d, \Lambda$ in the metric, which obey
a single algebraic constraint
\begin{align}
  c_0^2=3c_1d+1.
\label{relation}
\end{align}
As long as $c_1\neq0$, \eqref{metric} will not be a solution of Einstein gravity.
On the other hand, when $c_1=0$, the metric \eqref{metric} becomes identical to the
Schwarzschild-(A)dS solution of Einstein gravity. In this case, $\Lambda$ plays the role of
a cosmological constant, but it arises purely as an integration constant and is not
put in by hand.

In this paper, we will be concentrated in some other exotic aspects of conformal gravity
associated with the solution \eqref{metric}. We will restrict ourselves to two particular
aspects, i.e. the extended phase space thermodynamics and the heat engines.
Extended phase space thermodynamics for black holes is an active area of study in these days.
Basically the idea is to include more extensive variables such as the cosmological constant
(which is identified as an effective pressure) and take the black hole mass as enthalpy
rather than internal energy \cite{KastorEtal:2009,D.Kubiznak}. For most familiar black hole solutions, various types of phase
transitions have been found \cite{Wei:2012ui,Poshteh:2013pba,Cai:2013qga,Altamirano:2013uqa,Xu:2013zea,
Zou:2013owa,Altamirano:2014tva,Zou:2014mha,Liu:2014gvf,Johnson:2014xza,
Johnson:2014pwa,Frassino:2014pha,Dolan:2014vba,Lee:2014tma,Frassino:2015oca,
Lan:2015bia,Xu:2015hba,Altamirano:2013ane,Gunasekaran:2012dq,Xu:2014tja,Johnson:2013dka,Sun:2016til,Xu:2017wvu}, but that associated with the black hole solution \eqref{metric}
of conformal gravity looks most exotic, in the sense that only phase transitions of the
zeroth order are present \cite{Xu:2014kwa}. In this paper, we will see that there is an extra
level of exoticness in the extended phase space thermodynamics associated with the
solution \eqref{metric}. First, the equation of states (EOS) is always branched, and one
can take only one branch at a time. Second, the thermodynamical volume that is conjugate to the
effective pressure (let us consider the case $\Lambda<0$)
\begin{align}
  P=-\frac{\Lambda}{8\pi} \label{pre}
\end{align}
is {\em bounded from above}. As a consequence, the black hole is always super-entropic in one
branch and can again be super-entropic in the other branch of the EOS.

Heat engine that takes black hole as working media is also an interesting subject of study. With
both pressure and volume being dynamical, one may extract mechanical work by using
$P\mathrm{d}V$ term. A heat engine is defined by a closed cycle in the $P-V$ plane
\cite{Johnson:2014yja}, just like in classical thermodynamics of ordinary matter. One can
extract mechanical work through this cycle. Denoting the input and output heat flows
as $Q_H$ and $Q_C$ respectively, the work done in one cycle will be $W=Q_H-Q_C$, which is
exactly the area of the cycle in the $P-V$ plane. The efficiency of the heat engine is defined as
$\eta=\frac{W}{Q_H}$. This heat engine has been
constructed in RN AdS black holes \cite{Johnson:2014yja}, Gauss Bonnet AdS black holes \cite{Johnson:2015ekr}, Born-Infeld AdS black holes
\cite{Johnson:2015fva}, STU black holes \cite{Caceres:2015vsa} and Horava-Lifshitz black holes
\cite{Sadeghi:2016xal}. Since the extended phase space thermodynamics of conformal gravity looks different from other static black holes, it is natural to consider the heat engine construction in
its $P-V$ plane.

If we begin with a pair of isotherms with temperature $T_H$ and $T_C$ to build a heat engine, we have many ways to connect them. There are
two kinds of simple choices. The first one is using adiabatic paths. Since there is no heat flow along the
adiabatic paths, the heat engine is reversible. This is Carnot engine. It possesses the maximal efficiency $1-\frac{T_C}{T_H}$. The other one is using isochoric paths, which is Stirling engine. For usual
static black hole solutions such as the RN AdS black hole \cite{Johnson:2014yja}, the
entropy $S$ and thermodynamical volume $V$ are both monotonic functions of the horizon
radius and are not independent. The heat capacity $C_V=T(\frac{\partial S}{\partial T})_V=0$,
which means adiabatic and isochoric paths coincide. The Carnot engine and Stirling engine are identical. We will show that this is not
the case for static black holes in conformal gravity, where the thermodynamical volume and entropy are indeed independent. It would be interesting to investigate the difference between Carnot engine and
Stirling engine in conformal gravity.

Another interesting type of cycle is
the rectangular cycle which consists of two isobars and two isochors.
In \cite{Johnson:2016pfa} Clifford Johnson stated that cycles of arbitrary shape can be built
by using infinite number of small rectangular cycles, and a triangular cycle is constructed as an example. In another recent paper \cite{Chakraborty:2016ssb}, the authors initiated a benchmarking scheme that allows for cross-comparison of efficiency of different black hole heat engines by using a circular cycle. In this paper we will also study the rectangular heat engine associated with the
static solution \eqref{metric} of conformal gravity and consider the variation of the
efficiencies for the heat engines working between two fixed highest and lowest
temperatures.

\section{Black hole thermodynamics in conformal gravity revisited}

The original study of black hole
thermodynamics in conformal gravity can be found in \cite{Lu:2012xu},
and the extended phase space description as well as the corresponding critical behavior
were studied in detail in \cite{Xu:2014kwa}. Before rushing into the extended phase space
thermodynamics, let us point out a big difference between the study to be made below and
that in \cite{Xu:2014kwa}, namely we will be using the thermodynamical volume $V$, rather than
the black hole radius $r_0$, as a dimension in the thermodynamical phase space.
This different choice of state parameter is required, because we will need such a description
in order to construct the associated heat engines.

Besides the effective pressure \eqref{pre}, let us present the other necessary thermodynamical
quantities, quoted from \cite{Lu:2012xu}: the conserved charge associated with the timelike
Killing vector is identified as the enthalpy $H$,
\begin{align}
  H=\,{\frac {\left( c_{{1}}c_{{0}}-c_{{1}}-16\pi P\,d
 \right) }{12\pi }}.
\end{align}
The temperature is identified with the surface gravity at the horizon with radius $r_0$,
\begin{align}
  T=\,{\frac {8\pi P\,r_0^{3}-3\,c_{{0}}r_{{0}}-6\,d}
  {12\pi \,{r_{{0}}}^{2}}},
\label{temperature}
\end{align}
where $r_0$ is the largest root of $f(r)=0$, and its conjugate, i.e. the entropy, is
\begin{align}
  S=\,{\frac {\,\left( r_0-c_{{0}}r_{{0}}-\,  3\,d \right) }{3r_{{0}}}}.
\label{S}
\end{align}
We will not set the parameter $\Xi\equiv c_1$ to zero, because this object can
be interpreted as a massive spin-2 hair \cite{Lu:2012xu} and it will be preferrable to
keep it nonzero. Then, the thermodynamical conjugate of $\Xi$ is given by
\begin{align}
  \Psi=\,{\frac {\, \left( c_{{0}}-1 \right) }{12\pi }}.
  \label{Psi}
\end{align}
And finally, the thermodynamic volume reads
\begin{align}
  V=\left(\frac{\partial H}{\partial P}\right)_{S,\Xi}
  =-\,{\frac {2d}{3 }}.
  \label{V}
\end{align}

Although the case with $c_1\neq0$ is not a solution of Einstein gravity,
it is still instructive to make a comparison to the familiar AdS black hole solution of
Einstein gravity by taking the limit $c_1\to 0$. Then it is clear that the parameter
$d$ (which has to be negative) must somehow be
related to the black hole mass, and $c_0$ should be of topological nature (which can take
only three discrete values $0,\pm 1$ in Einstein gravity). Restoring the parameter $c_1$
the constraint \eqref{relation} can be solved by taking\footnote{Mathematically, there is
nothing wrong to solve $c_1$ in terms of $c_0$ and $d$, as did in \cite{Lu:2012xu}.
However, from the thermodynamical perspective, it is preferable to take $c_1$, rather
than $c_0$, as an independent thermodynamical parameter.}
\begin{align}
c_0=\pm\sqrt{3c_1d+1}.
\label{c0}
\end{align}
Now $c_0$ can take continuous values, but its signature can never be changed by
continuously varying the dependent parameters $c_1$ and $d$.
Consequently, when considering the extended phase space
thermodynamics, the signature of $c_0$ must be pre-assigned, and any process which changes
the signature of $c_0$ is thermodynamically prohibited.

With the above thermodynamical variables, we can now describe the first law of black hole
thermodynamics,
\begin{align}
  {\it \mathrm{d}H}=T{\it \mathrm{d}S}+ \Psi\,
  \mathrm{d}\Xi + V\,\mathrm{d}P.
\end{align}
The scaling dimensions of the above thermodynamical parameters/functions can be easily
determined, yielding $[S]=[\Psi]=[\mathrm{length}]^0,\,[H]=[T]=[\Xi]=[\mathrm{length}]^{-1},\,
[V]=[\mathrm{length}],\, [P]=[\mathrm{length}]^{-2}$. The Smarr formula then follows from the
homogeneity of $H$ \cite{Lu:2012xu},
\begin{align}
  H=2PV+\Xi \Psi\label{Smarr}.
\end{align}

Notice that here the parameter $\alpha$ in the gravitational action has been set equal to 1 for fixing the unit system. Since $\alpha$ does not change the equations of motion, the metric \eqref{metric} is independent of it. However, some thermodynamical quantities, such as enthalpy $H$ and black hole entropy $S$, should include contributions from the action. In fact, both $H$ and $S$ are proportional to $\alpha$($H\rightarrow\alpha H, S\rightarrow \alpha S$) \cite{Lu:2012xu}. Since $P$ and $\Xi$ remain unchanged, we know the $\Psi$ and $V$ are also proportional to $\alpha$ from \eqref{Psi} and \eqref{V}. Fortunately, this will not change the Smarr formula \eqref{Smarr}. See \cite{Kastor:2010gq} for more discussion about the Smarr formula.

For the purpose of studying heat engines in conformal
gravity, we need to begin with the EOS on the $P-V$ plane, fixing the
other extensive parameter $\Xi$. In this paper, we will restrict ourselves to the particular
choice $\Xi=1$. The EOS
arises from the expression \eqref{temperature} for the temperature $T$.
Using the condition $f(r_0)=0$ we can get
\begin{align}
  c_0=-\frac{8}{3}\pi Pr_0^2-r_0-\frac{d}{r_0}.
  \label{fr0}
\end{align}
Inserting \eqref{fr0} into eq.(\ref{temperature}) we have
\begin{align}
  &T=\frac{4}{3}Pr_0+\frac{1}{4\pi}-\frac{d}{4\pi r_0^2}\label{eq2}.
\end{align}
In the above two equations, $c_0$ is dependent on $d$ via eq.\eqref{c0} and $d=-3V/2$,
thanks to eq.\eqref{V}. So, it remains to eliminate $r_0$ using \eqref{fr0} and substitute
the result in \eqref{eq2} to get the final EOS. However, since eq.\eqref{fr0} is cubic in
$r_0$ for which the solution is quite complicated, it is preferable not to solve $r_0$
explicitly. Instead, we take $r_0$ as an intermediate parameter and present the EOS
as a pair of parametric equations.

Depending on the sign of $c_0$, the parametric EOS takes different forms, so it is necessary to
present them differently.

\subsubsection*{1) The case of $c_0<0$}

In this case \eqref{fr0} becomes
\begin{align}
-\sqrt{1+3d}=-\frac{8}{3}\pi Pr_0^2-r_0-\frac{d}{r_0}
\label{eq1}
\end{align}
Solving eq.\eqref{eq2} and \eqref{eq1} and making use of \eqref{V}, we get two
solutions for $P(T,r_0)$ and $V(T,r_0)$.
One of them reads
\begin{align}
  &P=\,{\frac {T}{2r_{{0}}}}-\,{\frac {r_{{0}}- \,\sqrt
{1-4\,\pi \,T r_0^{2}}}{8{\pi }
{r_{{0}}}^{2}}},
  &V=\frac{2r_0}{9}\bigg(4\pi Tr_0-r_0-2\sqrt{1-4\pi Tr_0^2}\bigg),
\label{solution1}
\end{align}
in which $r_0$ takes value in the range $\frac{2}{1+4\pi T}\leq r_0\leq
\frac{1}{\sqrt{4\pi T}}$, and the other branch reads
\begin{align}
  &P=\,{\frac {T}{2r_{{0}}}}-\,{\frac {r_{{0}}+ \,\sqrt
{1-4\,\pi T r_0^{2}}}{8{\pi }
{r_{{0}}}^{2}}},
  &V=\frac{2r_0}{9}\bigg(4\pi Tr_0-r_0+2\sqrt{1-4\pi  Tr_0^2}\bigg),
\label{solution2}
\end{align}
in which $\frac{1}{\sqrt{1+4\pi T}}<r_0<\frac{1}{\sqrt{4\pi T}}$. The upper and lower bounds for
the parameter $r_0$ are determined by the requirement that $P$ and $V$ should be both real and
positive.

Note that the two solutions \eqref{solution1} and \eqref{solution2} correspond to the same $c_0$
and meanwhile, they actually join together in a smooth way along any isotherm. The necessity
of subdividing the isotherm into the above two segments is just because the parametric equations
\eqref{solution1} and \eqref{solution2} may yield different values of $P$ and $V$ for the same
$r_0$. Nevertheless, let us keep in mind that in an isothermal process, the system may pass
smoothly from the EOS \eqref{solution1} to \eqref{solution2} or vice versa.

\subsubsection*{2) The case of $c_0>0$}
In this case the \eqref{fr0} becomes
\begin{align}
\sqrt{1+3d}=-\frac{8}{3}\pi Pr_0^2-r_0-\frac{d}{r_0}
\label{eq3}
\end{align}
Solving eq.\eqref{eq2} and \eqref{eq3} we obtain only one branch of solution
\begin{align}
  &P=\,{\frac {T}{2r_{{0}}}}-\,{\frac {r_{{0}}+ \,\sqrt
{1-4\,\pi T r_0^{2}}}{8{\pi }
{r_{{0}}}^{2}}},
  &V=\frac{2r_0}{9}\bigg(4\pi Tr_0-r_0+2\sqrt{1-4\pi  Tr_0^2}\bigg),
\label{solution3}
\end{align}
where $\frac{1}{\sqrt{16 \pi^2 T^2-4\pi T+1}}< r_0<\frac{1}{\sqrt{1+4\pi T}}$ so that $P$ is
real and positive.

\begin{figure}
\begin{center}
\includegraphics[width=0.5\textwidth]{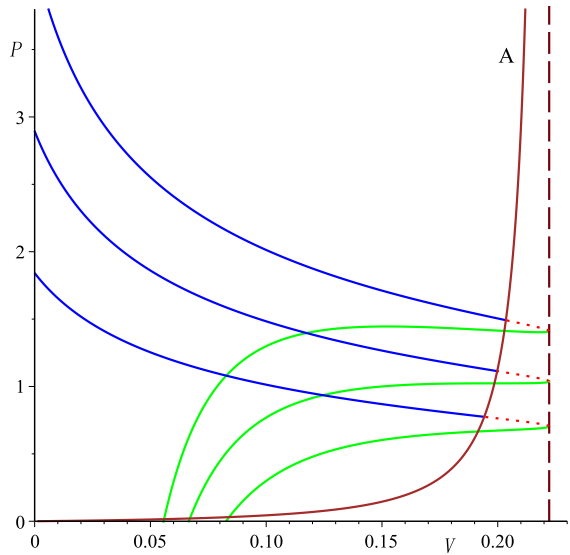}
\caption{Isothermal plots of the EOS \eqref{solution1}, \eqref{solution2} and \eqref{solution3}.
The blue lines, red dotted lines and green lines correspond to \eqref{solution1},
\eqref{solution2} and \eqref{solution3} respectively. From top to bottom the temperature
decreases. The vertical dashed line represents $V=\frac{2}{9}$ or $c_0=0$. Line A corresponds to
$r_0=\frac{1}{\sqrt{4\pi T}}$.}
\label{fig0}
\end{center}
\end{figure}

Though the equations in \eqref{solution3} take the same form as in \eqref{solution2},
they correspond to different sign choices for $c_0$, and therefore \eqref{solution2} and
\eqref{solution3} must be considered to be belonging to different branches of solutions.

In Fig.\ref{fig0} we present the isothermal plots of the EOS. Each isotherm can be divided into
two branches, i.e. the upper branch and the lower one. In the upper branch, each isotherm
is further subdivided into two segments, with the curve marked $A$ denoting the boundary
between the two segments. The vertical slashed line corresponds to $c_0=0$ or $V=2/9$,
which is the maximal thermodynamical volume that the black holes can take at any permissible
temperatures. It can be clearly seen that one can only take one
branch of solution at a time, otherwise isotherms at different temperatures could intersect
each other (which is physically implausible). The criticalities studied in \cite{Xu:2014kwa}
are associated with the lower branch \eqref{solution3} with $c_0>0$. In this paper, however,
in order to take the black hole as a stable working media for the heat engines, we will be
concentrated in the case $c_0<0$.

The above discussions have revealed that the thermodynamics for the spherical black holes in
conformal gravity is quite exotic in that the isotherms are branched and that the thermodynamic
volume is bounded from above. Now let us have a further look at the exoticness by examining the
so-called reverse isoperimetric inequality (RII).  The thermodynamical volume for black holes
is conjectured to satisfy the RII \cite{CveticEtal:2011}
\begin{align}
R\equiv \Big(\frac{(D-1)V}{{\cal{A}}_{D-2}}\Big)^{\frac{1}{D-1}}
\Big(\frac{{\cal{A}}_{D-2}}{A}\Big)^{\frac{1}{D-2}}\geq 1,
\label{Inequality}
\end{align}
where $A$ is the horizon area, and ${\cal{A}}_{D-2}$ is the area of a unit $(D-2)$-dimensional
sphere. This inequality indicates that the spherical boundary
minimizes surface area for a given volume. However, some recent works revealed that there exist
some black holes which violate the RII, examples of such black holes include the rotating AdS
black hole in ultra-spinning limit \cite{Hennigar:2014cfa} and Lifshitz black holes
\cite{Brenna:2015pqa}. These black holes are known as super-entropic.

In conformal gravity, the thermodynamical volume is not a monotonic function of the horizon
radius. It depends not only on $r_0$ but also on the temperature $T$. This is the principal
difference between the spherical solution \eqref{metric} for conformal gravity and
those for other theories of gravity. Now let us check whether the RII holds in conformal
gravity. In our case, $D=4$, $A=4\pi r_0^2$ and ${\cal{A}}_{2}=4\pi$. Inserting these
values into \eqref{Inequality} and using the expressions of $V$ in \eqref{solution1}, we get
\begin{align}
R&=\Big[\frac{1}{6\pi r_0^2}\Big(4\pi Tr_0-r_0-2\sqrt{1-4\pi Tr_0^2}\Big)\Big]^{1/3}.
\label{R1}
\end{align}
It is easy to find that
the equation $R=0$ in this case always has a real positive solution $r_0=\frac{2}{1+4\pi T}$,
which indicates the breakdown of the RII.
If the expressions of $V$ in \eqref{solution2} or \eqref{solution3} are used instead, we
would have, either
\begin{align}
R=\Big[\frac{1}{6\pi r_0^2}\Big(4\pi Tr_0-r_0+2\sqrt{1-4\pi Tr_0^2}\Big)\Big]^{1/3},
\frac{1}{\sqrt{1+4\pi T}}<r_0<\frac{1}{\sqrt{4\pi T}}
\label{R2}
\end{align}
or
\begin{align}
R=\Big[\frac{1}{6\pi r_0^2}\Big(4\pi Tr_0-r_0+2\sqrt{1-4\pi Tr_0^2}\Big)\Big]^{1/3},
\frac{1}{\sqrt{16 \pi^2 T^2-4\pi T+1}}<r_0<\frac{1}{\sqrt{1+4\pi T}}.
\label{R3}
\end{align}
Eq. \eqref{R2}, which arises from using \eqref{solution2}, still
corresponds to $c_0<0$. The break down of the RII in this branch has already been argued above.
Eq. \eqref{R3} corresponds to $c_0>0$. In this branch, we may evaluate $R$ at the
largest possible value of $r_0$, i.e. $r_0=1/\sqrt{1+4\pi T}$, yielding
\begin{align}
R_0=\bigg(\frac{1}{6\pi}\bigg)^{\frac{1}{3}}\sqrt{1+4\pi T}. \label{R0}
\end{align}
It can be shown that $R$ is monotonically decreasing with $r_0$.
Therefore, if $R_0$ in \eqref{R0} obeys $R_0\geq1$, then we will have $R\geq1$ for any
allowed $r_0$. A straightforward numeric check indicates that when $T> 0.4840$, we will
indeed have $R_0>1$. Therefore, in the branch $c_0>0$, the RII can be holding when $T> 0.4840$
and it may be broken otherwise.

In conclusion, the solution \eqref{metric} for conformal gravity can be super-entropic
in the branch $c_0>0$ in certain range of temperatures and is always super-entropic in the
branch $c_0<0$.

\section{Heat engines in the upper branch}

As advocated earlier, what we would like to do is to study the heat engine associated with the
extended phase space thermodynamics of the black hole solution \eqref{metric}. Effectively
we would take the black hole as working media, analyze the efficiencies for several kinds
of heat engines and study various impact factors that can affect the working efficiencies.
For this purposes it is preferable to consider only the upper branch of the EOS.
Moreover, to avoid the overwhelming complexities that may be caused by passing from the
parametric EOS \eqref{solution1} to \eqref{solution2}, we will consider only the heat engines
whose working cycles resides solely to the left of the curve $A$ as shown in Fig.\ref{fig0}.
The types of heat engines to be considered include Carnot, Stirling and
rectangular engines.

\subsection{Carnot engine}

Each heat engine corresponds to a cyclic process in the $P-V$ plane. For the Carnot engine, the
cyclic process is consisted of two isothermal and two adiabatic segments.
Since the heat flow does not take place along the adiabatic paths, the heat engine is
reversible.

To actually construct a Carnot engine in conformal gravity, we need to identify the equation
for the adiabats. According to \eqref{c0}, we have
\begin{align}
c_0=-\sqrt{1+3d}. \label{c0d}
\end{align}
Inserting the above equation and \eqref{V} into \eqref{S}, the entropy becomes
\begin{align}
S=\frac{1}{6}\bigg(\sqrt{4-18 V}+\frac{9V}{r_0}+2\bigg). \label{SV}
\end{align}
Adiabatic process obeys $\mathrm{d}S=0$. Using \eqref{SV} one can get the
following differential equation
\begin{align}
\frac{\mathrm{d}V}{\mathrm{d}r_0}=\frac{V \sqrt{4-18 V}}{r_0(\sqrt{4-18 V}-r_0)},
\end{align}
from which it follows that in the adiabatic process $r_0$ and $V$ are linked to each other via
\begin{align}
r_0=\frac{9V}{C+\sqrt{4-18 V}},
\end{align}
where $C$ is an integration constant. Inserting this result into \eqref{eq1} and replacing
$d$ with $V$ using \eqref{V}, we get
the equation for the adiabats
\begin{align}
P=\frac{1}{1296\pi V^2}\Big(C^3-12C-2(9V+4)\sqrt{4-18 V}\Big).
\label{adiabat}
\end{align}
Finally, using \eqref{S}, \eqref{V} and \eqref{adiabat}, we find that the entropy along each
adiabat is characterized solely by the constant $C$,
\begin{align}
S=\frac{1}{6}(C+2).
\end{align}
Clearly, the non-negativity of the entropy requires $C\geq -2$.

\begin{figure}[!htbp]
\begin{center}
\includegraphics[width=0.4\textwidth]{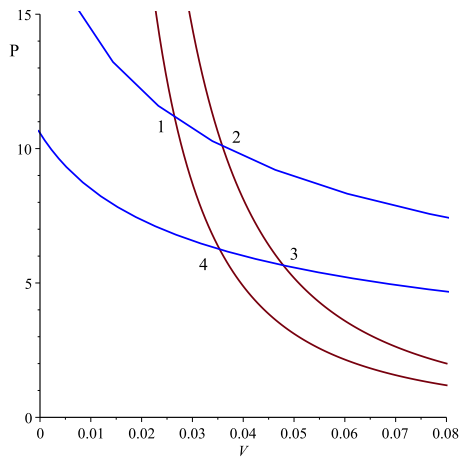}
\caption{Carnot cycle: $1\rightarrow2$ and $4\rightarrow3$ represents two isotherms
at with temperatures $T_1=T_2=2.0000$ and $T_4=T_3=1.5000$ respectively,
$1\rightarrow4$ and $2\rightarrow3$ are two adiabats corresponding to $C_1=C_4=4.0000$ and
$C_2=C_3=4.5000$ respectively.}
\label{fig2}
\end{center}
\end{figure}

In Fig.\ref{fig2} and Table \ref{tab1}, we give an example for the Carnot engine. The
isotherms $1\rightarrow2$ and $4\rightarrow3$ correspond to $T=2.0000$ and $T=1.5000$,
and the adiabats $1\rightarrow4$ and $2\rightarrow3$ correspond to $C=4.0000$ and $C=4.5000$,
respectively. Using \eqref{solution1} and \eqref{adiabat}, we can calculate the heat transferred
at each stage of the cyclic process $1 \to 2 \to 3\to4\to1$ numerically. The heat transferred
along the isotherms $1\rightarrow2$ and $3\rightarrow4$ are given respectively as
\begin{align}
Q_{1\rightarrow2}&=T_{1}(S_2-S_1),\\
Q_{3\rightarrow4}&=T_{4}(S_4-S_3).
\end{align}
On each adiabats, the entropy is fixed, so $S_1=S_4$ and $S_2=S_3$, the efficiency becomes
\begin{align}
\eta=1-\frac{-Q_{3\rightarrow4}}{Q_{1\rightarrow2}}=1-\frac{T_4}{T_1},
\end{align}
which depends only on the two working temperatures, as expected. The thermodynamical
parameters and functions of states at each corner of the Carnot cycle are listed in
Table \ref{tab1}.

\begin{table}[!htbp]
\centering
\begin{tabular}{|c|c|c|c|c|c|}
\hline

Point ~&~  $P$ ~&~   $V$ ~&~ $T$  ~&~   $S$ ~&~   $r_0$  \\
\hline
$1$ ~&~ $11.1390$ ~&~ $0.0265$ ~&~ $2.0000$ ~&~ $1.0000$ ~&~ $0.1124$ \\
\hline
$2$ ~&~ $10.0810$ ~&~ $0.0359$ ~&~ $2.0000$ ~&~ $1.0833$ ~&~ $0.1211$ \\
\hline
$3$ ~&~ $5.6560$ ~&~ $0.0479$ ~&~ $1.5000$ ~&~ $1.0833$ ~&~ $0.1580$ \\
\hline
$4$ ~&~ $6.2516$ ~&~ $0.0354$ ~&~ $1.5000$ ~&~ $1.0000$  ~&~ $0.1470$ \\
\hline
\end{tabular}\label{tab2}
\caption{Thermodynamical parameters and functions of states for the Carnot cycle
described in Fig.\ref{fig2}.}
\label{tab1}
\end{table}

\subsection{Stirling Heat engine}

Unlike the Carnot cycle which consists of two isotherms and two adiabats, the Stirling
cycle is consisted of two isotherms and two isochors. For usual static black holes in Einstein
gravity, the entropy depends solely on the thermodynamical volume $V$, hence the adiabats
are also isochors, and Carnot engines are simultaneously Stirling eingines.
However, in conformal gravity, $T$ and $V$ are independent variables and the entropy
depends also on both of them, as can be seen in eq.\eqref{SV} ($r_0$ in \eqref{SV} can be
seen as a function of $V$ and $T$, thanks to the EOS \eqref{solution1}).
This is among the major differences between the extended phase space thermodynamics for
conformal gravity and that for Einstein gravity.

During a Stirling cycle, the heat transfer also takes place along the isochoric paths.
It is essential to calculate the heat $Q = \int T\mathrm{d} S$ at
constant $V$. In order to do so, we need to write the temperature $T$ and the entropy $S$ as
functions of $V$ and $r_0$.

Inserting \eqref{eq1} and \eqref{V} into \eqref{temperature}, we obtain the temperature as a function of $V$ and $r_0$
\begin{align}
T=T(V,r_0)=\frac{2r_0\left(\sqrt{4-18V}-r_0\right)+9V}{8\pi r_0^2}.
\end{align}
Similarly the entropy can be rewritten as
\begin{align}
S=S(V,r_0)=\frac{1}{6}\left(\sqrt{4-18V}+2+\frac{9V}{r_0}\right).
\end{align}
Then the heat transferred at constant $V$ is the integral of the following equation,
\begin{align}
T\mathrm{d}S|_V&=\frac{3}{16}\frac{(2r_0^2-2\sqrt{4-18V}r_0-9V)V}{\pi r_0^4}\mathrm{d}r_0.
\label{TdSV}
\end{align}

\begin{figure}[!htbp]
\begin{center}
\includegraphics[width=0.45\textwidth]{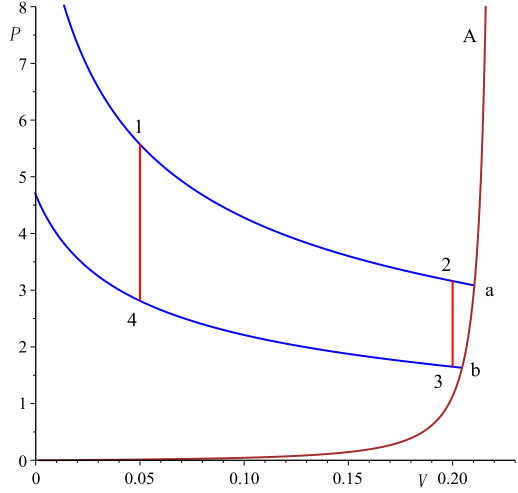}
\caption{Stirling cycle: the two isotherms $1\rightarrow2$, $4\rightarrow3$ are taken at
$T_1=T_2=1.5000$ and $T_3=T_4=1.0000$ and the two isochors $1\rightarrow4$, $2\rightarrow3$
are taken at $V_1=0.0500$ and $V_2=0.2000$ respectively. For simplicity,
we design the cycle so that the marked points $2$ and $3$ are located to the left of the curve
$A$.}
\label{fig3}
\end{center}
\end{figure}

In Fig.\ref{fig3} and Table \ref{tab2} we give an example of Stirling engine, where
$T_1=T_2=1.50$, $T_4=T_3=1.00$, and $V_1=V_4=0.05$, $V_2=V_3=0.20$. In order to calculate the
heat transferred during the isochoric process, we need to integrate \eqref{TdSV} from one
temperature (say, $T_1$) to another (e.g. $T_4$), keeping $V$ fixed. For this to be done,
we need to re-express $r_0$ as a function of $T$ and $V$ using the EOS \eqref{solution1}.
This can be done analytically, but the expressions are not worth quoting.
We will directly apply a numerical method to calculate the heat. All the
thermodynamic parameters and functions of states are listed in Table \ref{tab2}.
On the isotherms $1\rightarrow2$ and $3\rightarrow4$, the heat transferred are given by
\begin{align}
Q_{1\rightarrow2}&=T_{1}(S_2-S_1)=0.9746,\\
Q_{3\rightarrow4}&=T_{4}(S_4-S_3)=-0.5174.
\end{align}
On the other hand, the heat transferred during $2\rightarrow3$ and $4\rightarrow2$ are,
respectively,
\begin{align}
Q_{2\rightarrow3}&=\int^{(r_{0})_3}_{(r_{0})_2}T\mathrm{d}S=-0.3022, \\
Q_{4\rightarrow1}&=\int^{(r_{0})_1}_{(r_{0})_4}T\mathrm{d}S=0.1380.
\end{align}
Here $(r_0)_k$ represents the value of $r_0$ at the marked point $k$, which is determined
via the EOS \eqref{solution1}.

\begin{table}[!htbp]
\centering
\begin{tabular}{|c|c|c|c|c|c|}
\hline

Point ~&~  $P$ ~&~   $V$ ~&~ $T$  ~&~   $S$ ~&~   $r_0$  \\
\hline
$1$ ~&~ $5.5720$ ~&~ $0.0500$ ~&~ $1.5000$ ~&~ $1.0965$ ~&~ $0.1597$ \\
\hline
$2$ ~&~ $3.1607$ ~&~ $0.2000$ ~&~ $1.5000$ ~&~ $1.7462$ ~&~ $0.2295$ \\
\hline
$3$ ~&~ $1.6495$ ~&~ $0.2000$ ~&~ $1.0000$ ~&~ $1.5028$ ~&~ $0.2819$ \\
\hline
$4$ ~&~ $2.8118$ ~&~ $0.0500$ ~&~ $1.0000$ ~&~ $0.9855$  ~&~ $0.2091$ \\
\hline
\end{tabular}\label{tab2}
\caption{Thermodynamic parameters and functions of states for the Stirling cycle
described in Fig. \ref{fig3}.}
\label{tab2}
\end{table}

The efficiency of the Stirling engine can be calculated as
\begin{align}
\eta=1-\frac{-Q_{2\rightarrow3}-Q_{3\rightarrow4}}{Q_{1\rightarrow2}+Q_{4\rightarrow1}}=0.2633.
\end{align}
For comparison, the Carnot engine working between the same two temperatures $T_1$ and $T_4$ has
the efficiency
\begin{align}
\eta_C=1-\frac{T_{4}}{T_{1}}=0.3333.
\end{align}
So we have
\begin{align}
\frac{\eta}{\eta_C}=0.7899,
\end{align}
which is consistent with the fact that the Carnot efficiency $\eta_C$ is the maximum efficiency
that any heat engine can have at the prescribed working temperatures $T_1$ and $T_4$.

\begin{figure}
\begin{center}
\includegraphics[width=0.45\textwidth]{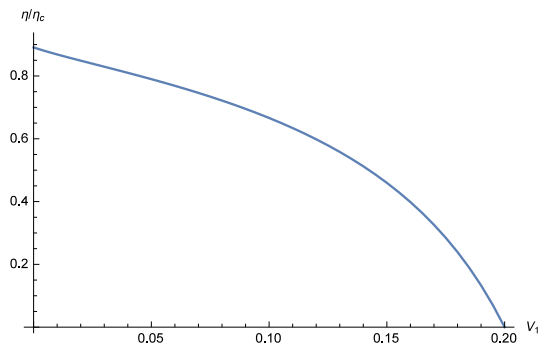}
\caption{Plot of $\eta/\eta_C$ as a function of $V_1$ at $T_1=1.5000$, $T_4=1.0000$ and $V_2=0.2000$.}
\label{fig4}
\end{center}
\end{figure}

Fixing the values of $T_1$, $T_4$ and $V_2$, we can explore the dependence of the Stirling
efficiency on $V_1$. In Fig.\ref{fig4} we present the plot of $\eta/\eta_C$ as a
function of $V_1$. We find that the ratio $\eta/\eta_C$ achieves its maximal value
${\eta}/{\eta_C}\simeq 0.8908$ as $V_1\rightarrow 0$. When $V_1$ becomes larger,
the efficiency decreases. It approaches $0$ when $V_1$
gets close to $V_2$. This is very different from the RN-AdS black holes in Einstein gravity,
for which the Stirling engine coincides with the Carnot engine, so that the efficiency is always
$\eta_C$ no matter how narrow the cycle is. In conformal gravity , the efficiency of Stirling engine
depends on the specific circle. When the cycle becomes narrow, the efficiency decreases.

\subsection{Rectangular heat engine}

Now we study the rectangular engine whose working cycle is composed of two isobars and
two isochors. On the $P-V$ plane, this working cycle is represented by a rectangle, thus
the name rectangular engine. The work done through this cycle is exactly the area of the
rectangle. It can be used as the seed of an algorithm to build heat engine with arbitrary shape.
In \cite{Johnson:2016pfa,Chakraborty:2016ssb}, a triangular cycle and circular cycle were presented as examples.

The heat transferred along the isochors has been investigated in the last subsection. It remains
to calculate heat transferred in isobaric processes. For this purposes, the temperature $T$ and
the entropy $S$ must be rewritten as functions of $P$ and $r_0$. This can be done by combined
use of \eqref{S}, \eqref{fr0}, \eqref{eq2} and \eqref{solution1}, however the result is again
branched. The first branch of solution is
\begin{align}
  &T=2Pr_0+\frac{1}{8\pi}-\frac{\sqrt{4-3r_0^2-32\pi P r_0^3}}{8\pi r_0},
  &S=\frac{1}{3}+\frac{8\pi}{3}Pr_0^2-\frac{1}{3}\sqrt{4-3r_0^2-32\pi P r_0^3},
\label{solution4}
\end{align}
where the black hole radius $r_0$ satisfies $\frac{2}{1+4\pi T} \leq r_0\leq \frac{2}{\sqrt{1+16\pi T}}$.
The second one is
\begin{align}
  &T=2Pr_0+\frac{1}{8\pi}+\frac{\sqrt{4-3r_0^2-32\pi P r_0^3}}{8\pi r_0},
  &S=\frac{1}{3}+\frac{8\pi}{3}Pr_0^2+\frac{1}{3}\sqrt{4-3r_0^2-32\pi P r_0^3},
\label{solution5}
\end{align}
where $r_0$ satisfies $\frac{2}{\sqrt{1+16\pi T}}<r_0<\frac{1}{\sqrt{4\pi T}}$.

For a rectangular heat engine, if we specify the temperature $T_2,T_4$ and the thermodynamic
volume $V_2,V_4$, the rectangle is uniquely determined, and the pressure $P_1=P_2$ and
$P_4=P_3$ can be computed directly.
A novel kind of complexities that is not seen in the former two kinds of heat engines
arises due to the above branching of solutions. During a working cycle, the rectangular engine
may/may not get across the border curve $B$ defined by the implicit function
$r_0=\frac{2}{\sqrt{1+16\pi T}}$ in either an isobaric or an isochoric process,
so, we need to consider several different cases.

\begin{figure}[!htbp]
\begin{center}
\includegraphics[width=0.55\textwidth,height=0.35\textheight]{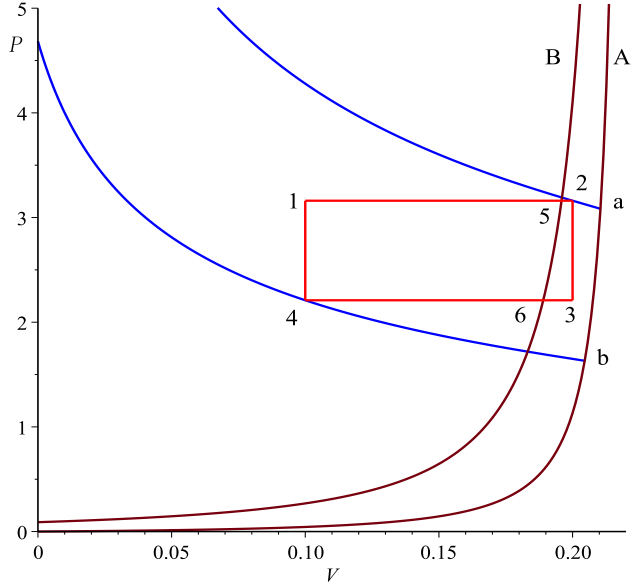}
\caption{An example of rectangular engine: the marked points 5 and 6 refer to the
intersection of isobars and the curve B. The state 2 corresponds to
$T_2=1.5000$, $V_2=0.2000$, and the state $4$ corresponds to $T_4=1.000$ and $V_4=0.1000$. }
\label{fig6}
\end{center}
\end{figure}

\begin{table}[!htbp]
\centering
\begin{tabular}{|c|c|c|c|c|c|}
\hline

Point ~&~  $P$ ~&~   $V$ ~&~ $T$  ~&~   $S$ ~&~   $r_0$  \\
\hline
$1$ ~&~ $3.1607$ ~&~ $0.1000$ ~&~ $1.2439$ ~&~ $1.2782$ ~&~ $0.2150$ \\
\hline
$2$ ~&~ $3.1607$ ~&~ $0.2000$ ~&~ $1.5000$ ~&~ $1.7462$ ~&~ $0.2295$ \\
\hline
$3$ ~&~ $2.2096$ ~&~ $0.2000$ ~&~ $1.1982$ ~&~ $1.6049$ ~&~ $0.2573$ \\
\hline
$4$ ~&~ $2.2096$ ~&~ $0.1000$ ~&~ $1.0000$ ~&~ $1.1932$  ~&~ $0.2448$ \\
\hline
$5$ ~&~ $3.1607$ ~&~ $0.1959$ ~&~ $1.4907$ ~&~ $1.7283$  ~&~ $0.2295$ \\
\hline
$6$ ~&~ $2.2096$ ~&~ $0.1890$ ~&~ $1.1785$ ~&~ $1.5625$  ~&~ $0.2577$ \\
\hline
\end{tabular}\label{tab3}
\caption{Thermodynamic parameters and functions of states for the rectanglar cycle
described in Fig.\ref{fig6}}
\label{tab3}
\end{table}

In Fig.\ref{fig6} and Table \ref{tab3}
we give an example for this kind of heat engine. Here we take $T_2=1.5000$, $T_4=1.0000$,
$V_2=0.2000$, and $V_4=0.1000$. $T_2$ and $T_4$ are respectively the lowest and highest
temperatures that the engine can attain during a working cycle.
Two isobars $1\rightarrow 5 \rightarrow2$, $4\rightarrow6 \rightarrow3$ and two isochors
$1\rightarrow4$, $2\rightarrow3$ constitute the whole rectanglar cycle. Using
\eqref{solution4}, the heat transferred in the isobaric processes $1\rightarrow5$ and
$6\rightarrow4$ are calculated to be
\begin{align*}
Q_{1\rightarrow5}&=\int_{1\to 5} T(P_1,r_0)\mathrm{d}S(P_1,r_0)=0.6166,\\
Q_{6\rightarrow4}&=\int_{6\to 4} T(P_4,r_0)\mathrm{d}S(P_4,r_0)=-0.4028.
\end{align*}
On the other hand, using \eqref{solution5}, the heat transferred in the processes
$5\rightarrow2$ and $3\rightarrow6$ are obtained, yielding
\begin{align*}
Q_{5\rightarrow2}&=\int_{5\to 2} T(P_1,r_0)\mathrm{d}S(P_1,r_0)=0.0267,\\
Q_{3\rightarrow6}&=\int_{3\to 6} T(P_4,r_0)\mathrm{d}S(P_4,r_0)=-0.0504.
\end{align*}
Moreover, the heat transferred during the isochoric processes $2\rightarrow3$ and
$4\rightarrow1$ are respectively
\begin{align*}
Q_{2\rightarrow3}&=\int_{2\to 3} T(V_2,r_0)\mathrm{d}S(V_2,r_0)=-0.1902,\\
Q_{4\rightarrow1}&=\int_{4\to 1} T(V_1,r_0)\mathrm{d}S(V_1,r_0)=0.0951.
\end{align*}
We can check that the total work done in one rectangular cycle is precisely the area of the
rectangle,
\begin{align*}
Q_{1\rightarrow5}+Q_{5\rightarrow2}+Q_{2\rightarrow3}+Q_{3\rightarrow6}+
Q_{6\rightarrow4}+Q_{4\rightarrow1}
=(P_1-P_4)(V_2-V_1)=0.0951.
\end{align*}
The efficiency of the rectangular engine is then given by
\begin{align*}
\eta=1-\frac{-Q_{2\rightarrow3}-Q_{3\rightarrow6}-Q_{6\rightarrow4}}
{Q_{1\rightarrow5}+Q_{5\rightarrow2}+Q_{4\rightarrow1}}=0.1288.
\end{align*}

The rectangular engine does not work at constant temperature at any stage of the cycle.
In order to make comparison between the efficiency of rectangular engines and that of the
Carnot engines, we take a Carnot engine which works between the temperatures $T_2$ and $T_4$
(the maximum and the minimum temperatures during a rectangular cycle),
which has the efficiency
\begin{align*}
\eta_C=1-\frac{T_{4}}{T_{2}}=0.3333.
\end{align*}
So we have
\begin{align*}
\frac{\eta}{\eta_C}=0.3863,
\end{align*}
indicating that the rectangular engine is much inefficient than the Carnot engine.

\begin{figure}
\begin{center}
\includegraphics[width=0.45\textwidth]{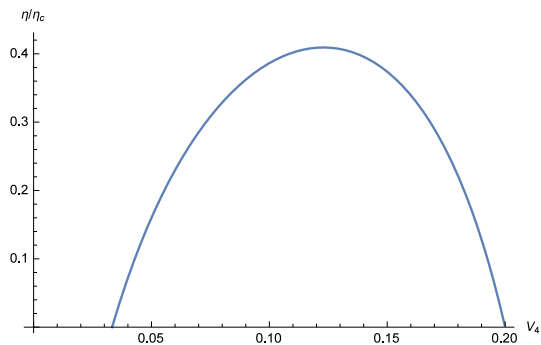}
\caption{Plot of ${\eta}/{\eta_C}$ as a function of $V_4$ at fixed $T_2=1.5000$, $V_2=0.2000$
and $T_4=1.0000$.}
\label{fig7}
\end{center}
\end{figure}

Now let us vary $V_4$ while keeping $T_2$, $V_2$ and $T_4$ fixed. In Fig.\ref{fig7} we depict
the dependence of the efficiency of
the rectangular engine on $V_4$. The ratio ${\eta}/{\eta_C}$ goes to zero at two distinct
points. The first one is at $V_4\simeq 0.0333$, which corresponds to $P_4\rightarrow P_1$.
The second one corresponds to $V_4\rightarrow V_2=0.2000$. In both cases, the rectangular
cycles become extremely narrow, which means the net work done during one cycle goes to zero.
Notice that in static spacetimes where the Carnot engine and Stirling engine coincide, the
efficiency of the rectangular engines approaches the Carnot efficiency at leading order as
$V_4\rightarrow V_2$. In conformal gravity, the efficiency goes to zero. This happens because
heat transfer occurs also on the isochors. There is an extreme value of ${\eta}/{\eta_C}$
which depends on the concrete choice of the engine.

\begin{figure}[!htbp]
\begin{center}
\includegraphics[width=0.45\textwidth]{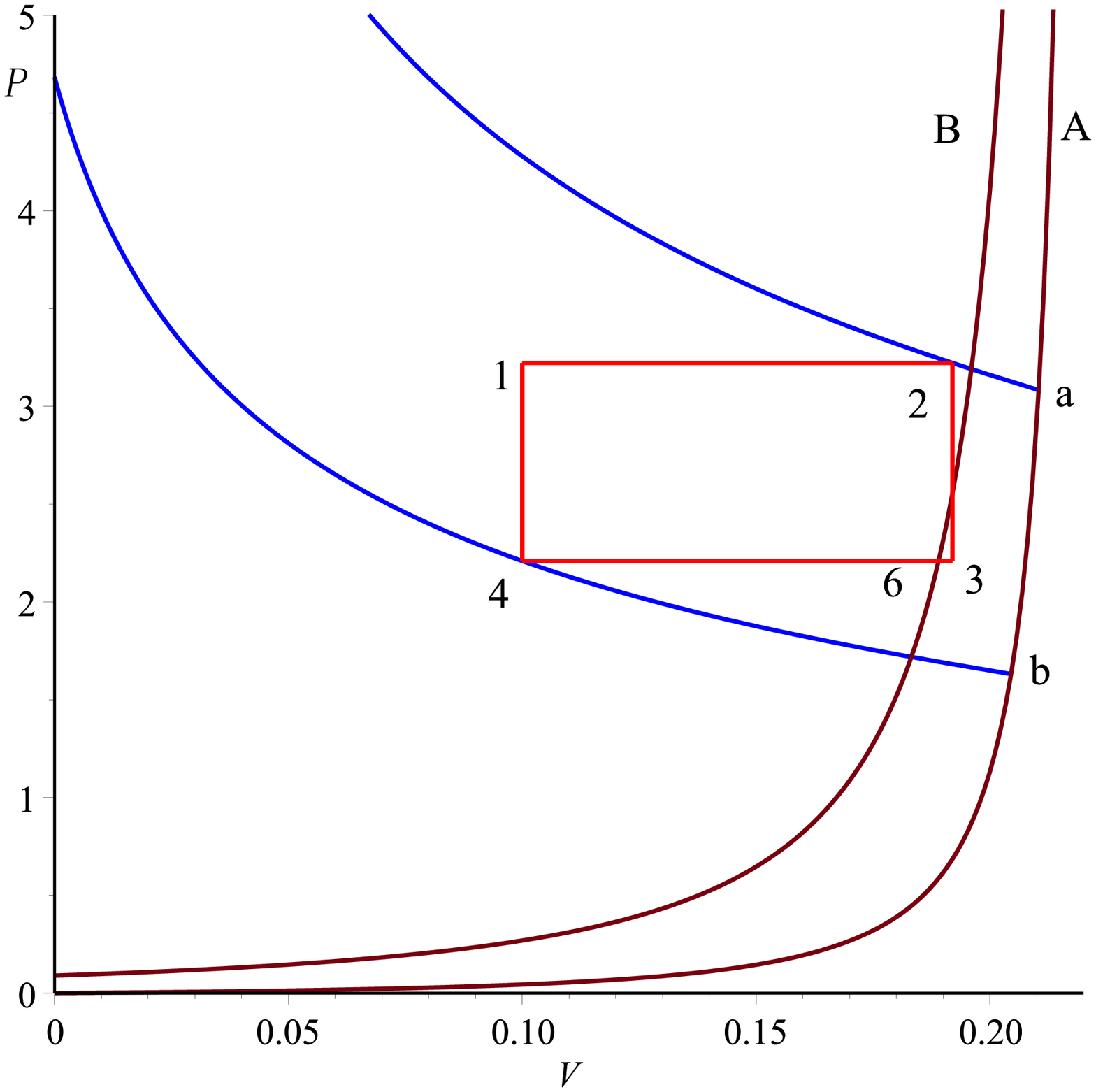}
\includegraphics[width=0.45\textwidth]{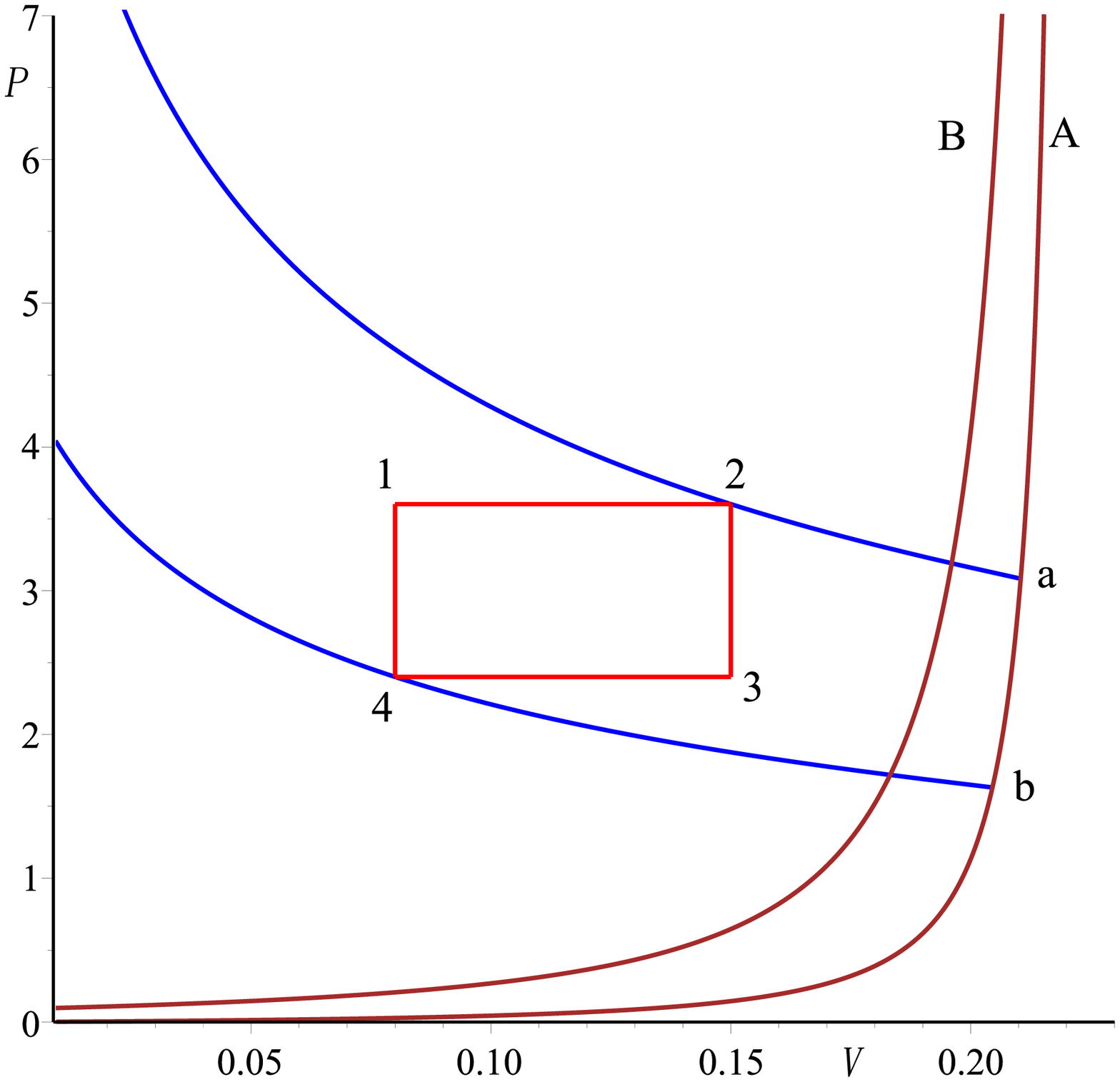}
\caption{Two more examples of rectangular engines. Left:  $T_2=1.5000$, $V_2=0.1920$,
$T_4=1.0000$ and $V_4=0.1000$. Right:
$T_2=1.5000$, $V_2=0.1500$, $T_4=1.0000$ and $V_4=0.0800$. }
\label{fig8}
\end{center}
\end{figure}

\begin{figure}[!htbp]
\begin{center}
\includegraphics[width=0.45\textwidth]{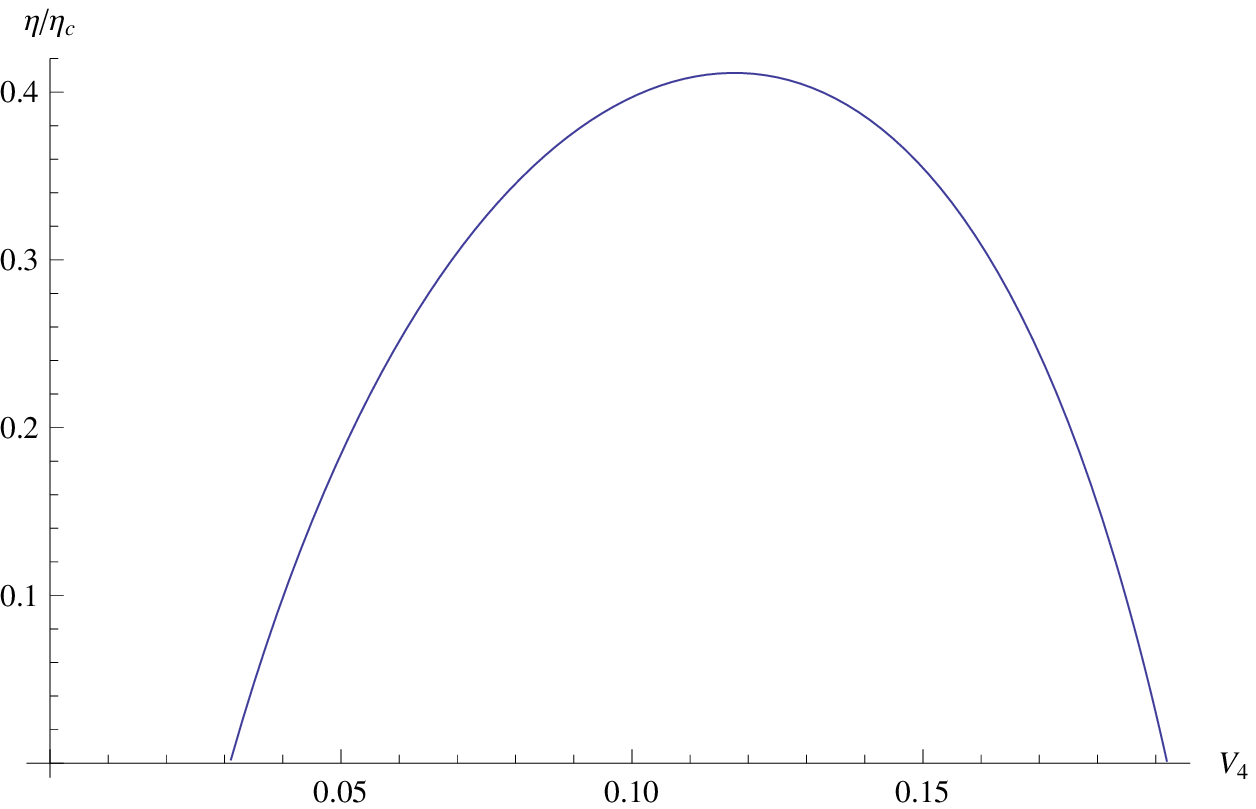}
\includegraphics[width=0.45\textwidth]{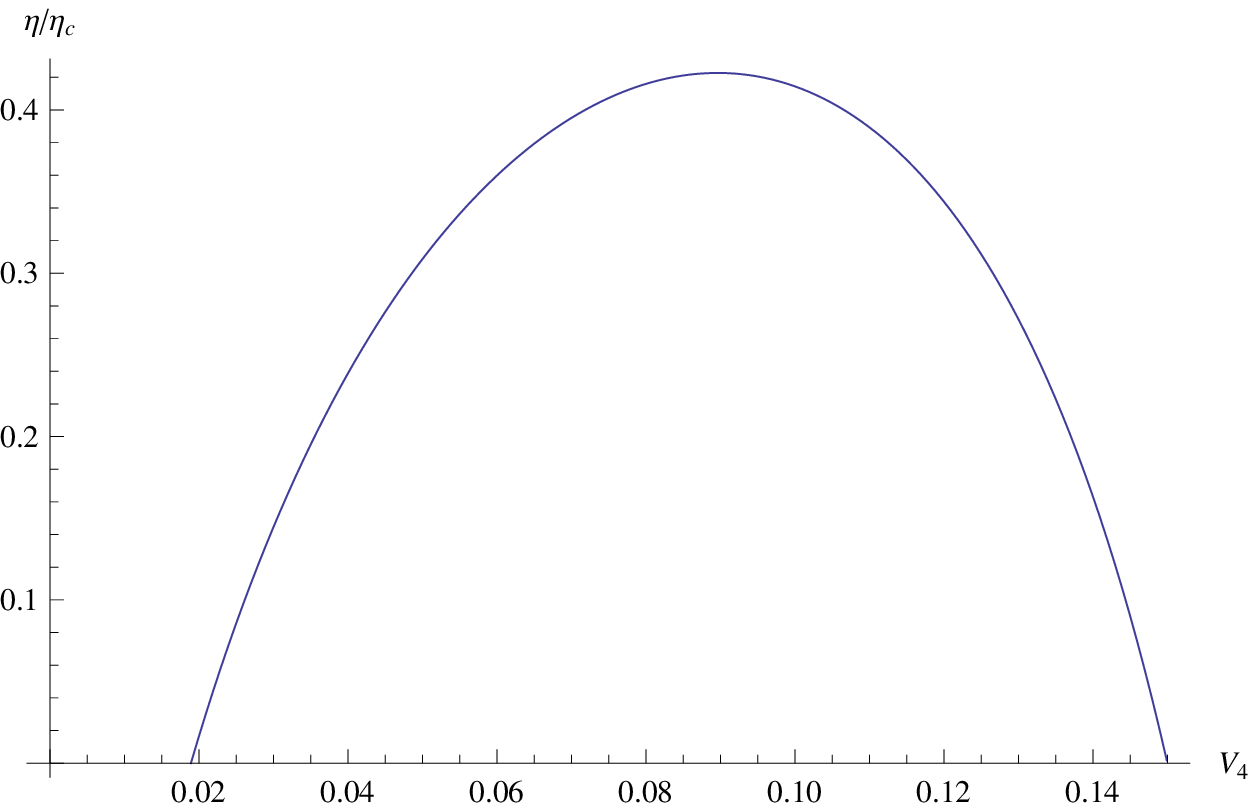}
\caption{Plot of ${\eta}/{\eta_C}$ as a function of $V_4$ at fixed $V_2$:
Left: $T_2=1.5000$, $V_2=0.1920$, $T_4=1.0000$; Right:  $T_2=1.5000$, $V_2=0.1500$, $T_4=1.0000$. }
\label{fig9}
\end{center}
\end{figure}

In Fig.\ref{fig8} the cases of two other possible choices of $V_2$ and $V_4$ are presented.
On the left plot, we set $V_2=0.1920$ and $V_4=0.1000$. In this case only the
isobaric process $3\to 6\to 4$ intersects with the curve $B$. Since the thermodynamical
volume $V$ does not suffer from the second branching represented by the curve $B$, we can
safely ignore the intersection between the isochoric process $2\to 3$ with the curve $B$.
On the right plot, we set $V_2=0.1500$ and $V_4=0.0800$. In this case the
whole rectangle is located to the left of the curve $B$, so there is no need to consider the
complexities arising from the second level of the branching. Fixing the states $2$ in both cases
and varying the state $4$ along the lower isotherm, one gets the plots of $\eta/\eta_C$
as functions of $V_4$. These are presented in  Fig.\ref{fig9}.
It can be seen that is all cases there is an maximum for ${\eta}/{\eta_C}$.

\section{Summary}

In this paper, we investigate the extended phase space thermodynamics and heat engine construction in four dimensional conformal gravity. Although conformal gravity is an attractive model of gravity in four dimensions, its thermodynamics is very different from Einstein gravity. The EOS for the AdS black hole in conformal gravity is always branched, and any process which drives the system from one branch to the other is thermodynamically prohibited. The thermodynamical volume is bounded from above, and the black hole is always super-entropic in one branch and can be super-entropic in the other branch at certain range of temperature. Because of these exotic behaviors, it is particularly interesting to see what happens if we build heat engines in the $P-V$ plane.

We consider three interesting kinds of heat engines, Carnot engine, Stirling engine and rectangle engine. For usual static black hole solutions, such as RN AdS black hole, the entropy depends solely on the thermodynamical volume $V$. Adiabatic and isochoric paths coincide, hence the Carnot engine and Stirling engine are identical. Since there is no heat flow along the isochoric paths, the efficiency of rectangle engine always approaches Carnot efficiency $\eta_C$ as $V_1\rightarrow V_2$. However, in conformal gravity, the $T$ and $V$ are independent variables and the entropy depends on both of them. Because heat flows also take place during the isochoric paths, the Carnot and Stirling
engines do not coincide. We obtain the adiabatic curves and build the Carnot engine in the $P-V$ plane, which gives exactly the $\eta_C$. For the Stirling engine, the efficiency is always smaller than $\eta_C$. When the circle of Stirling engine becomes narrow, the efficiency decreases. For the rectangle engine, if we make the cycle extremely narrow, the efficiency approaches zero. There is an extreme value of ${\eta}/{\eta_C}$ which depends on the concrete choice of the engine.

\textbf{Notes added.} Recently, the authors of \cite{Hennigar:2017apu} developed a new method to calculate the heat flows on the rectangular heat engine. Instead of calculating $\int T\mathrm{d}S$ directly, they applied the first law of black hole thermodynamics. The heat flow on the isobar $1\rightarrow 2$ can be obtained as
\begin{align}
 Q_{1\to 2}&=\int_{1\to 2} T\mathrm{d}S=\int_{1\to 2}\mathrm{d}H-\int_{1\to 2}V\mathrm{d}P=H_2-H_1,
\end{align}
while the heat flow on the isochore $2\rightarrow 3$ is
\begin{align}
 Q_{2\to 3}&=\int_{2\to 3} T\mathrm{d}S=\int_{2\to 3}\mathrm{d}(H-PV)+\int_{2\to 3}P\mathrm{d}V=H_3-H_2-V_2(P_3-P_2),
\end{align}
where $H_k$ represents the value of enthalpy $H$ at the marked point $k$. We can also calculate the heat flows on $3\rightarrow 4$ and $4\rightarrow1$ similarly, hence the efficiency $\eta$ can be obtained. For example, in the rectangular heat engine of Fig.\ref{fig6}, we have $H_1=0.5859$, $H_2=1.2294$, $H_3=0.8489$ and $H_4=0.3957$. One can check the methods in \cite{Hennigar:2017apu} and current paper give the same result.

\section*{Acknowledgment}
We would like to thank the referee whose suggestions and comments
largely helped us in improving the original manuscript. This work is supported by the National Natural Science Foundation of China under the grant No. 11575088.


\begin{thebibliography}{100}


\bibitem{Maldacena:2011mk}
  J.~Maldacena,
  ``Einstein Gravity from Conformal Gravity,''
    [\eprint{1105.5632}].

\bibitem{Mannheim:1988dj}
  P.~D.~Mannheim and D.~Kazanas,
  ``Exact Vacuum Solution to Conformal Weyl Gravity and Galactic Rotation Curves,''
  Astrophys.\ J.\  {\bf 342}, 635 (1989).

\bibitem{Mannheim:2005bfa}
  P.~D.~Mannheim,
  ``Alternatives to dark matter and dark energy,''
  Prog.\ Part.\ Nucl.\ Phys.\  {\bf 56}, 340 (2006)
        [\eprint{astro-ph/0505266}].



\bibitem{Mannheim:2010ti}
  P.~D.~Mannheim and J.~G.~O'Brien,
  ``Impact of a global quadratic potential on galactic rotation curves,''
  Phys.\ Rev.\ Lett.\  {\bf 106}, 121101 (2011)
  [\eprint{1007.0970}].



\bibitem{Mannheim:2011ds}
  P.~D.~Mannheim,
  ``Making the Case for Conformal Gravity,''
  Found.\ Phys.\  {\bf 42}, 388 (2012)
  [\eprint{1101.2186}].


\bibitem{Riegert:1984zz}
  R.~J.~Riegert,
  ``Birkhoff's Theorem in Conformal Gravity,''
  Phys.\ Rev.\ Lett.\  {\bf 53}, 315 (1984).


\bibitem{KastorEtal:2009}
 D.~Kastor, S.~Ray and J.~Traschen,
 ``Enthalpy and the Mechanics of AdS Black Holes,''
  Class.\ Quant.\ Grav.\  {\bf 26}, 195011 (2009)
  [\eprint{0904.2765}].

\bibitem{D.Kubiznak}
 D.~Kubiznak and R.~B.~Mann,
 ``P-V criticality of charged AdS black holes,''
  JHEP {\bf 1207}, 033 (2012)
  [\eprint{1205.0559}].




\bibitem{Wei:2012ui}
 S.~-W.~Wei and Y.~-X.~Liu,
  ``Critical phenomena and thermodynamic geometry of charged Gauss-Bonnet AdS black holes,''
  Phys.\ Rev.\ D {\bf 87}, no. 4, 044014 (2013)
  [\eprint{1209.1707}].


\bibitem{Poshteh:2013pba}
  M.~B.~J.~Poshteh, B.~Mirza and Z.~Sherkatghanad,
  ``Phase transition, critical behavior, and critical exponents of Myers-Perry black holes,''
  Phys.\  Rev.\ D {\bf 88}, 024005 (2013)
  [\eprint{1306.4516}].

\bibitem{Cai:2013qga}
  R.~-G.~Cai, L.~-M.~Cao, L.~Li and R.~-Q.~Yang,
  ``P-V criticality in the extended phase space of Gauss-Bonnet black holes in AdS space,''
  JHEP {\bf 1309}, 005 (2013)
  [\eprint{1306.6233}].




\bibitem{Altamirano:2013uqa}
  N.~Altamirano, D.~Kubiznak, R.~B.~Mann and Z.~Sherkatghanad,
  ``Kerr-AdS analogue of tricritical point and solid/liquid/gas phase transition,''
  [\eprint{1308.2672}].

\bibitem{Xu:2013zea}
  W.~Xu, H.~Xu and L.~Zhao,
 ``Gauss-Bonnet coupling constant as a free thermodynamical variable and the associated criticality,''
  Eur.\ Phys.\ J.\ C {\bf 74}, 2970 (2014)
  [\eprint{1311.3053}].


\bibitem{Zou:2013owa}
  D.~-C.~Zou, S.~-J.~Zhang and B.~Wang,
  ``Critical behavior of Born-Infeld AdS black holes in the extended phase space thermodynamics,''
  Phys.\ Rev.\ D {\bf 89}, 044002 (2014)
  [\eprint{1311.7299}].




\bibitem{Altamirano:2014tva}
  N.~Altamirano, D.~Kubiznak, R.~B.~Mann and Z.~Sherkatghanad,
  ``Thermodynamics of rotating black holes and black rings: phase transitions and thermodynamic volume,''
   Class.\ Quant.\ Grav.\  {\bf 31} (2014) 042001
  [\eprint{1401.2586}].


\bibitem{Zou:2014mha}
  D.~-C.~Zou, Y.~Liu and B.~Wang,
  ``Critical behavior of charged Gauss-Bonnet AdS black holes in the grand canonical ensemble,''
    Phys.\ Rev.\ D {\bf 90}, no. 4, 044063 (2014)
   [\eprint{1404.5194}].

\bibitem{Liu:2014gvf}
  Y.~Liu, D.~-C.~Zou and B.~Wang,
  ``Signature of the Van der Waals like small-large charged AdS black hole phase transition in quasinormal modes,''
   JHEP {\bf 1409}, 179 (2014)
  [\eprint{hep-th/1405.2644}].

\bibitem{Johnson:2014xza}
  C.~V.~Johnson,
  ``Thermodynamic Volumes for AdS-Taub-NUT and AdS-Taub-Bolt,''
  Class.\ Quant.\ Grav.\  {\bf 31}, no. 23, 235003 (2014)
  [\eprint{1405.5941}].

\bibitem{Johnson:2014pwa}
  C.~V.~Johnson,
  ``The Extended Thermodynamic Phase Structure of Taub-NUT and Taub-Bolt,''
  Class.\ Quant.\ Grav.\  {\bf 31}, 225005 (2014)
  [\eprint{1406.4533}].


\bibitem{Frassino:2014pha}
  A.~M.~Frassino, D.~Kubiznak, R.~B.~Mann and F.~Simovic,
 ``Multiple Reentrant Phase Transitions and Triple Points in Lovelock Thermodynamics,''
  JHEP {\bf 1409}, 080 (2014)
  [\eprint{1406.7015}].

\bibitem{Dolan:2014vba}
  B.~P.~Dolan, A.~Kostouki, D.~Kubiznak and R.~B.~Mann,
 ``Isolated critical point from Lovelock gravity,''
  Class.\ Quant.\ Grav.\  {\bf 31}, no. 24, 242001 (2014)
    [\eprint{1407.4783}].

\bibitem{Lee:2014tma}
  C.~O.~Lee,
  ``The extended thermodynamic properties of Taub每NUT/Bolt每AdS spaces,''
  Phys.\ Lett.\ B {\bf 738}, 294 (2014)
  [\eprint{1408.2073}].


\bibitem{Frassino:2015oca}
  A.~M.~Frassino, R.~B.~Mann and J.~R.~Mureika,
  ``Lower-Dimensional Black Hole Chemistry,''
 [\eprint{1509.05481}].

\bibitem{Lan:2015bia}
  S.~Q.~Lan, J.~X.~Mo and W.~B.~Liu,
  ``A note on Maxwell＊s equal area law for black hole phase transition,''
  Eur.\ Phys.\ J.\ C {\bf 75}, no. 9, 419 (2015)
    [\eprint{gr-qc/1503.07658}].


\bibitem{Xu:2015hba}
  H.~Xu and Z.~M.~Xu,
  ``Maxwell's equal area law for Lovelock Thermodynamics,''
  Int.\ J.\ Mod.\ Phys.\ D {\bf 26}, 1750037 (2017)
  doi:10.1142/S0218271817500377
  [\eprint{1510.06557}].


\bibitem{Altamirano:2013ane}
  N.~Altamirano, D.~Kubiznak and R.~B.~Mann,
  ``Reentrant Phase Transitions in Rotating AdS Black Holes,''
  Phys.\ Rev.\ D {\bf 88}, 101502 (2013)
  [\eprint{1306.5756}].



\bibitem{Gunasekaran:2012dq}
  S.~Gunasekaran, R.~B.~Mann and D.~Kubiznak,
  ``Extended phase space thermodynamics for charged and rotating black holes and Born-Infeld vacuum polarization,''
  JHEP {\bf 1211}, 110 (2012)
  [\eprint{1208.6251}].



\bibitem{Xu:2014tja}
  H.~Xu, W.~Xu and L.~Zhao,
 ``Extended phase space thermodynamics for third order Lovelock black holes in diverse dimensions,''
  Eur.\ Phys.\ J.\ C {\bf 74}, no. 9, 3074 (2014)
  [\eprint{1405.4143}].



\bibitem{Johnson:2013dka}
  C.~V.~Johnson,
  ``Large N Phase Transitions, Finite Volume, and Entanglement Entropy,''
  JHEP {\bf 1403}, 047 (2014)
    [\eprint{1306.4955}].

\bibitem{Sun:2016til}
  Y.~Sun, H.~Xu and L.~Zhao,
  ``Thermodynamics and holographic entanglement entropy for spherical black holes in 5D Gauss-Bonnet gravity,''
  JHEP {\bf 1609}, 060 (2016)
 [\eprint{1606.06531}].

\bibitem{Xu:2017wvu}
  H.~Xu,
  ``Entanglement growth during Van der Waals like phase transition,''
  [\eprint{1705.02604}].

\bibitem{Xu:2014kwa}
  W.~Xu and L.~Zhao,
  ``Critical phenomena of static charged AdS black holes in conformal gravity,''
  Phys.\ Lett.\ B {\bf 736}, 214 (2014)
    [\eprint{1405.7665}].

\bibitem{Johnson:2014yja}
  C.~V.~Johnson,
  ``Holographic Heat Engines,''
  Class.\ Quant.\ Grav.\  {\bf 31}, 205002 (2014)
     [\eprint{1404.5982}].



\bibitem{Johnson:2015ekr}
  C.~V.~Johnson,
  ``Gauss-Bonnet Black Holes and Holographic Heat Engines Beyond Large N,''
    [\eprint{1511.08782}].


\bibitem{Johnson:2015fva}
  C.~V.~Johnson,
  ``Born Infeld AdS black holes as heat engines,''
  Class.\ Quant.\ Grav.\  {\bf 33}, no. 13, 135001 (2016)
      [\eprint{1512.01746}].

\bibitem{Caceres:2015vsa}
  E.~Caceres, P.~H.~Nguyen and J.~F.~Pedraza,
  ``Holographic entanglement entropy and the extended phase structure of STU black holes,''
  JHEP {\bf 1509}, 184 (2015)
  [\eprint{1507.06069}].

\bibitem{Sadeghi:2016xal}
  J.~Sadeghi and K.~Jafarzade,
  ``The correction of Horava-Lifshitz black hole from holographic engine,''
        [\eprint{1604.02973}].

\bibitem{Johnson:2016pfa}
  C.~V.~Johnson,
  ``An Exact Efficiency Formula for Holographic Heat Engines,''
  Entropy {\bf 18}, 120 (2016)
  [\eprint{1602.02838}].

\bibitem{Chakraborty:2016ssb}
  A.~Chakraborty and C.~V.~Johnson,
  ``Benchmarking Black Hole Heat Engines,''
    [\eprint{1612.09272}].


\bibitem{Lu:2012xu}
  H.~Lu, Y.~Pang, C.~N.~Pope and J.~F.~Vazquez-Poritz,
  ``AdS and Lifshitz Black Holes in Conformal and Einstein-Weyl Gravities,''
  Phys.\ Rev.\ D {\bf 86}, 044011 (2012)
    [\eprint{1204.1062}].

\bibitem{Kastor:2010gq}
  D.~Kastor, S.~Ray and J.~Traschen,
  ``Smarr Formula and an Extended First Law for Lovelock Gravity,''
  Class.\ Quant.\ Grav.\  {\bf 27}, 235014 (2010)
    [\eprint{1005.5053}].

\bibitem{CveticEtal:2011}
  M.~Cvetic, G.~W.~Gibbons, D.~Kubiznak and C.~N.~Pope,
 ``Black Hole Enthalpy and an Entropy Inequality for the Thermodynamic Volume,''
  Phys.\ Rev.\ D {\bf 84}, 024037 (2011)
  [\eprint{1012.2888}].

\bibitem{Hennigar:2014cfa}
  R.~A.~Hennigar, D.~Kubiz防芍k and R.~B.~Mann,
  ``Entropy Inequality Violations from Ultraspinning Black Holes,''
  Phys.\ Rev.\ Lett.\  {\bf 115}, no. 3, 031101 (2015)
    [\eprint{1411.4309}].


\bibitem{Brenna:2015pqa}
  W.~G.~Brenna, R.~B.~Mann and M.~Park,
  ``Mass and Thermodynamic Volume in Lifshitz Spacetimes,''
  Phys.\ Rev.\ D {\bf 92}, no. 4, 044015 (2015)
    [\eprint{1505.06331}].

\bibitem{Hennigar:2017apu}
  R.~A.~Hennigar, F.~McCarthy, A.~Ballon and R.~B.~Mann,
  ``Holographic heat engines: general considerations and rotating black holes,''
  [\eprint{1704.02314}].





\end{thebibliography}

\providecommand{\href}[2]{#2}\begingroup
\footnotesize\itemsep=0pt
\providecommand{\eprint}[2][]{\href{http://arxiv.org/abs/#2}{arXiv:#2}}

\end{document}